# Sectoral contributions to sustainable development in Türkiye: Which sector is more effective?

**Emre Akusta**[*]

[*]*Ph.D., Assoc. Prof., Kırklareli University, Faculty of Economics and Administrative Sciences, Department of Economics, Kırklareli, 39100 Türkiye. E-mail: emre.akusta@klu.edu.tr. ORCID: https://orcid.org/0000-0002-6147-5443*

**ABSTRACT**

Enhancing sustainable development performance requires an assessment of the relative roles of economic sectors in this process. However, comparative empirical evidence regarding the sectoral structure of sustainable development is limited, particularly for Türkiye. Therefore, this study examines the long-run relationship between sectoral structure and sustainable development in Türkiye by focusing on agriculture, industry, construction, and services. The empirical analysis uses annual data for the period 2000–2022 and proceeds in three steps. First, the stationarity properties of the variables are examined using ADF, PP, and Zivot–Andrews unit root tests. The Johansen cointegration test is then applied to determine whether a long-run equilibrium relationship exists among the variables. Finally, long-run coefficients are estimated using the DOLS estimator, while the FMOLS estimator is used as a robustness check. The findings show that all sectoral shares are positively associated with the sustainable development index in the long run. Based on the DOLS results, the services sector has the highest coefficient at 0.882, followed by the agriculture, industry, and construction sectors with coefficients of 0.302, 0.265, and 0.193, respectively. The FMOLS robustness check supports the DOLS long-run estimates. The study contributes to the literature by providing comparative sector-level evidence for Türkiye. It also highlights that sustainable development strategies should not rely on a single sector, but should be designed through balanced and sector-specific policies that account for sectoral differences.

***Suggested citation:*** Akusta, E. (2026). Sectoral contributions to sustainable development in Türkiye: Which sector is more effective? Gazi Journal of Economics and Business, 12(2), 401-422. https://doi.org/10.30855/gjeb.2026.12.2.014

Doi: https://doi.org/10.30855/gjeb.2026.12.2.014

e-ISSN: 2548-0162

# Türkiye'de sürdürülebilir kalkınmaya sektörel katkılar: Hangi sektör daha etkili?

**ÖZ**

Sürdürülebilir kalkınma performansının güçlendirilmesi, ekonomik sektörlerin bu süreçteki göreli rollerinin anlaşılmasını gerektirmektedir. Ancak sürdürülebilir kalkınmanın sektörel yapısına ilişkin karşılaştırmalı ampirik kanıtlar, özellikle Türkiye için sınırlı düzeydedir. Bu nedenle bu çalışma, tarım, sanayi, inşaat ve hizmetler sektörlerine odaklanarak Türkiye'de sektörel yapı ile sürdürülebilir kalkınma arasındaki uzun dönemli ilişkiyi incelemektedir. Çalışmanın ampirik analizi 2000–2022 dönemine ait yıllık verilere dayanmakta ve üç aşamalı bir yöntem izlemektedir. İlk olarak, değişkenlerin durağanlık özellikleri ADF, PP ve Zivot–Andrews birim kök testleriyle incelenmiştir. Daha sonra, değişkenler arasında uzun dönemli bir denge ilişkisinin bulunup bulunmadığını belirlemek amacıyla Johansen eşbütünleşme testi uygulanmıştır. Son olarak, uzun dönem katsayıları DOLS tahmincisiyle tahmin edilmiş; FMOLS tahmincisi ise sağlamlık kontrolü olarak kullanılmıştır. Bulgular, tüm sektör paylarının uzun dönemde sürdürülebilir kalkınma endeksi ile pozitif ilişkili olduğunu göstermektedir. DOLS sonuçlarına göre, hizmetler sektörü 0,882 ile en yüksek katsayıya sahiptir; bunu sırasıyla 0,302, 0,265 ve 0,193 katsayılarıyla tarım, sanayi ve inşaat sektörleri izlemektedir. FMOLS sağlamlık kontrolü, DOLS uzun dönem tahminlerini desteklemektedir. Çalışma, Türkiye'de sürdürülebilir kalkınmanın sektörel yapısına ilişkin karşılaştırmalı ampirik kanıt sunarak literatüre katkı sağlamaktadır. Ayrıca sürdürülebilir kalkınma stratejilerinin tek bir sektöre dayanmaması; sektörlerin farklı rollerini dikkate alan dengeli ve sektör odaklı politikalarla tasarlanması gerektiğini vurgulamaktadır.

## 1. Introduction

The Industrial Revolution accelerated economic growth and increased the welfare of societies. However, this process has also brought environmental and natural resource problems. The increase in production and consumption has caused environmental problems that were initially perceived as local issues but later acquired a cross-border character (Pisani, 2006). While increasing industrialization and globalization have been the key to economic development, they have also brought global problems such as depletion of natural resources, deforestation and environmental pollution. These problems strengthened the idea that environmental balance should be protected in the second half of the 20th century. As a result, sustainable development emerged (Akusta, 2024).

The report Our Common Future provided the widely used definition of sustainable development (World Commission on Environment and Development [WCED], 1987). This definition emphasizes that economic growth alone is not sufficient and that environmental and social balances should also be considered. It also draws attention to intergenerational equity as a fundamental principle. The concept of basic needs implies the coexistence of a sound environment, a just society, and a well-functioning economy (Diesendorf, 2020). It is also emphasized that the justice to be achieved at the global level should be integrated with intergenerational justice. In this framework, economic development should be planned to meet the needs of not only current generations but also future generations. In this respect, sustainable development represents a holistic approach covering economic, environmental and social dimensions. The 17 Sustainable Development Goals adopted by the United Nations in 2015 are the most concrete indicator of this approach. These goals aim to provide comprehensive solutions in many areas such as combating climate change, transition to clean energy, sustainable production and consumption (Sarkar, 2013). However, achieving these goals requires countries not only to focus on economic growth, but also to take concrete steps on issues such as the environment and social equality.

Today, the rapid growth of the global economy has led to deepening environmental and social problems (Sachs, 2019). Current growth models are rapidly depleting natural resources and increasing environmental problems. This highlights the importance of balancing economic growth with environmental protection. Sustainable development refers to a development model that encompasses not only economic growth but also social equality and environmental protection. The development dynamics and economic structure of each country differ. In this regard, sectoral components of economic growth play a guiding role in determining development policies. Moreover, the sectoral structure of each country is quite different.

Türkiye's economic growth is shaped by key sectors such as agriculture, industry, services and construction. The agricultural sector is traditionally importance in Türkiye and plays a critical role in rural development and food security. The industrial sector is the engine of economic development. It also plays an important role in technological development and foreign trade. The service sector supports the economy, particularly through tourism and trade activities. The construction sector attracts attention with its infrastructure projects and employment generation capacity. However, the contributions of these sectors to sustainable development differ not only in terms of economic growth but also in terms of environmental and social impacts. Sustainable development in Türkiye is critical for both sustaining economic growth and managing environmental pressures. Increasing industrialization, energy consumption, and urbanization intensify environmental pressures. Innovative policies and sustainable development strategies are therefore needed to manage these pressures (Türkmen and Kılıç, 2020). Türkiye's sectoral sustainable development policy will be decisive in achieving sustainable development goals. Therefore, this study aims to analyze sectoral contributions to sustainable development in Türkiye and identify which sectors contribute more to this process.

This study contributes to the literature in five main ways: (1) To the best of our knowledge, empirical evidence that jointly examines the contributions of the main economic sectors to sustainable development within a unified country-level framework remains limited. Although previous studies have examined sectoral efficiency, sector-specific SDG impacts, and the sustainability implications of individual sectors, the comparative role of agriculture, industry, construction, and services in Türkiye’s sustainable development has not been sufficiently addressed. This study analyzes the long-run association between different sectors and sustainable development in Türkiye. Thus, this study aims to fill an important gap in the literature and examine the sectoral dimension of sustainable development policies in depth. (2) The Dynamic Ordinary Least Squares (DOLS) method was used in the analysis. Additionally, a robustness check was performed using the Fully Modified Ordinary Least Squares (FMOLS) method. This set of methods provides a powerful tool for estimating long-term relationships among cointegrated variables. (3) The study examines not only long-run relationships but also coefficient estimates of these relationships. This approach allows for a more detailed analysis of the size and direction of the long-run associations between sectoral shares and sustainable development. (4) The study uses the most recent available data for Türkiye covering the period 2000–2022. (5) The results of the study provide important implications for policymakers on the role of sectors in achieving sustainable development goals. In particular, the findings provide guidance for setting policy priorities on a sectoral basis.

The rest of the study is organized as follows. Section 2 presents the sectoral outlook in Türkiye. Section 3 reviews the relevant literature. Section 4 describes the data and methodology. Section 5 reports the empirical findings. Section 6 provides the FMOLS robustness check. Section 7 discusses the findings in relation to the literature and policy implications. Section 8 presents the results and policy recommendations.

## 2. Sectoral outlook in Türkiye

Türkiye's economic development process is shaped by the contributions of different sectors. The agriculture, industry, construction, and service sectors play a critical role in the country's sustainable development. Each sector may affect sustainable development through different economic, social, and environmental channels. However, these impacts differ in terms of their shares in GDP, their production capacities and their contribution to sustainable development. This section provides an overview of Türkiye's sectoral structure. The shares of agriculture, industry, construction, and services in GDP are compared over time to provide a descriptive basis for evaluating their role in sustainable development.

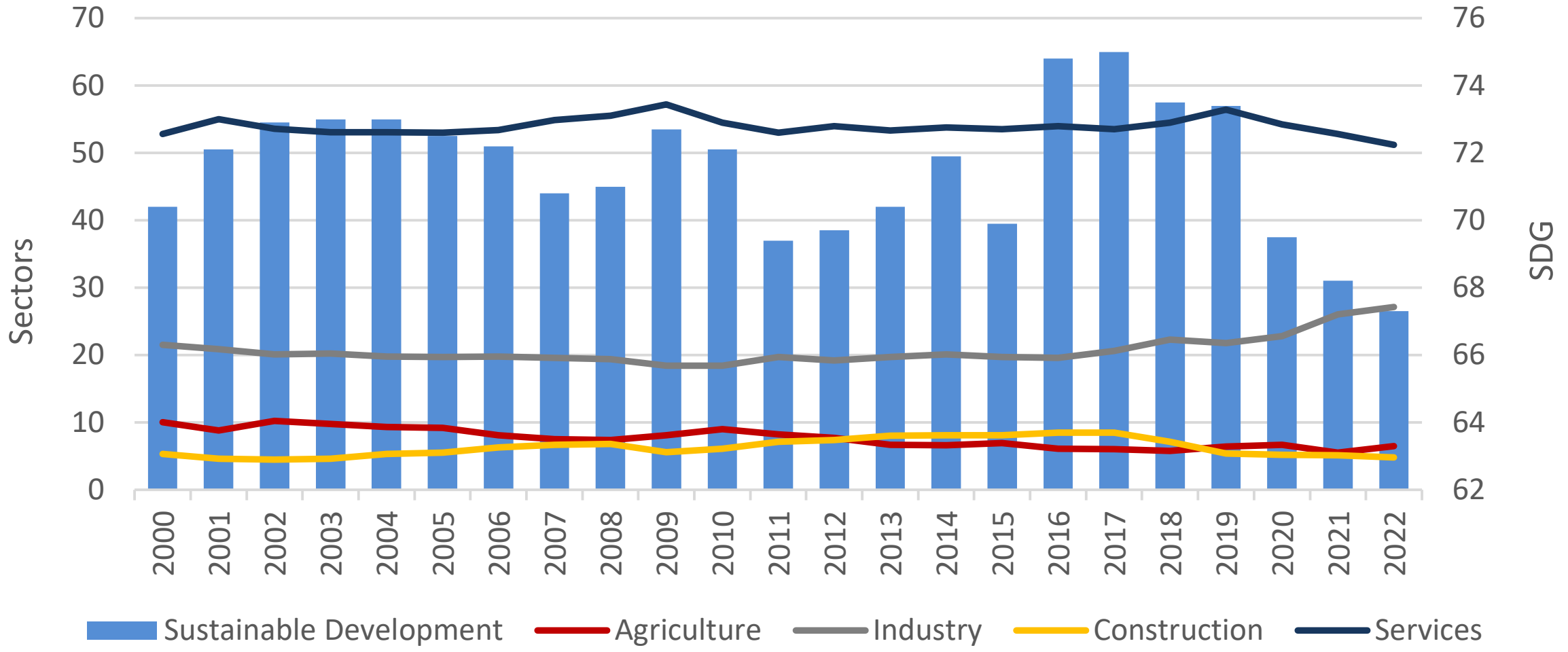

*Figure 1.* Sectors and sustainable development indicators for Türkiye.

Source: Constructed by the author using TurkStat data.

Figure 1 presents data on sectoral size and sustainable development in Türkiye. Türkiye's sustainable development index has shown a stable increase over the years. The index value was 70.4 in 2000 and gradually increased in the following years, indicating progress toward sustainable development (Sachs, Lafortune and Fuller, 2024). However, this progress has been shaped by the differential impact of economic activity across sectors.

An analysis of the share of the agricultural sector in GDP shows that it was around 10% in 2000, but gradually declined in the following years. By 2004, this share had fallen to 9.3% (MTF, 2024). Despite the traditional importance of the agricultural sector, its impact on economic growth is gradually declining. This decline suggests that the relative economic weight of agriculture has weakened, although the sector remains important for rural development and food security. The industrial sector can be defined as one of the engines of development. However, the share of industry in GDP is also declining. Data show that the share declined from 21.5% in 2000 to 19.8% in 2004 (MTF, 2024). This decline shows that the weight of the industrial sector in the economy is decreasing. However, the industrial sector still has an important place in the development process with its potential to create high added value.

The share of the construction sector in GDP remained at a relatively low level. While this ratio was 5.3% in 2000, it declined in 2001 and 2002. However, it rose again to 5.3% in 2004 (TR Ministry of Treasury and Finance [MTF], 2024). While the construction sector contributes to economic growth, it also draws attention with its environmental costs in terms of sustainable development. Therefore, when evaluating the sector's impact on the development process, not only its economic contributions but also its environmental consequences should be considered. The service sector stands out as the sector with the largest share in the Turkish economy. While the share of the service sector in GDP was 52.8% in 2000, it reached 55% in 2001 (MTF, 2024). This shows that the service sector has become the engine of economic growth. The sector makes significant contributions to sustainable development through the income and employment opportunities it generates, especially in sub-sectors such as tourism and trade.

The overall sectoral outlook reveals that the services sector has become increasingly prominent in the Turkish economy, while the weight of agriculture and industry in the economy has declined over time. Increasing the economic contribution of the services sector is important for sustainable development. However, it is critical for a balanced development model that the industrial and agricultural sectors are not neglected in this process.

## 3. Literature review

Sustainable development has become one of the most critical issues of the present century with its economic, environmental and social dimensions. This concept aims to produce solutions to global problems such as the rapid depletion of natural resources, climate change and social inequalities (Harris and Roach, 2017). The United Nations' Sustainable Development Goals have increased global awareness in this field. Thus, the number of studies addressing sustainable development in a multidimensional way has increased in the literature.

Recent studies examine sustainable development in different dimensions. One of these dimensions is environmental sustainability. For example, Alagöz (2007) emphasized the measures to be taken for environmental sustainability. The study stated that policies should be implemented to protect the ecological base. Similarly, Abdiraimov (2016) emphasized the importance of green energy use in sustainable development. It is suggested that private sector investments should be encouraged in this field. Karabıçak and Özdemir (2015) evaluated the impact of natural resource use on sustainable development and argued that environmentally friendly industrial structures should be developed.

The importance of renewable energy resources in environmental sustainability has been frequently emphasized in the literature. In Türkiye-specific studies, the contribution of renewable energy resources to environmental and economic development has been discussed in detail. For example, Koçaslan (2010) examined the potential of wind energy in Türkiye and revealed the positive impact of this resource on sustainable development. Acaravcı and Erdoğan (2018) showed that renewable energy generation reduces environmental pollution and promotes economic growth. Sadorsky (2009) stated that economic growth is a factor that increases green energy consumption. Sharif, Baris-Tuzemen, Uzuner, Ozturk and Sinha (2020) examined the impacts of renewable energy consumption on Türkiye's ecological footprint and emphasized the importance of green energy use in increasing environmental sustainability. Furthermore, Ecevit, Çetin and Yücel (2022) assessed the positive impacts of green energy consumption on public health. It was revealed that green energy provides health and social benefits beyond economic dimensions. In another important study on green energy, Yücel, Özdemir and Ayaz (2021) analyzed the role of green energy production incentives in promoting sustainable economic development. Cheng, Zhang, Kirschen, Huang and Kang (2020) determined that cost reductions have affected renewable energy consumption in China. Akram, Chen, Khalid, Ye and Majeed (2020) obtained similar results for developing countries, while Liu, Wang, Wang, Wang, Chang, He and Li (2020) confirmed these findings for the United Kingdom.

Another dimension of sustainable development is economic sustainability. Economic sustainability focuses on green economic policies, efficient use of natural resources, and the quality of production structures. In this context, Akusta (2026) shows that economic complexity improves environmental performance in BRICS-T countries, while economic growth, energy intensity, and population density negatively affect environmental performance. Özer (2016) found that Türkiye cannot fully utilize its green energy potential. It is stated that if this potential is utilized effectively, energy dependency can be reduced and economic competitiveness can be increased. Özaslan (2023) analyzed the contribution of green growth strategies in cities to economic development. He revealed that these strategies play a critical role in increasing environmental and economic sustainability. In the industrial sector, Akdağ and İskenderoğlu (2018) examined the effects of renewable energy consumption on economic growth in European Union member countries. The study emphasized the positive impacts of green energy use on economic growth. Sadorsky (2009) provides evidence that economic growth increases renewable energy consumption in developing countries. These findings show the strategic importance of renewable energy sources for economic development. Studies focusing on the relationship between economic growth and carbon emissions have been evaluated within the framework of the Environmental Kuznets Curve hypothesis. Lean and Smyth (2010) found unidirectional causality from carbon emissions to economic growth in five ASEAN countries. Similarly, Farhani and Rejeb (2012) found a unidirectional relationship from economic growth to carbon emissions in MENA countries. From the Turkish perspective, Bekar (2018) confirmed the impact of carbon emissions on economic growth.

Social sustainability is the third and final dimension of sustainable development. Social sustainability includes components such as education, health and social equality. Çelik (2006) emphasized the importance of health services in sustainable development and stated that investments in health would increase social welfare. Similarly, Yeni (2014) discusses the social dimensions of sustainable development and how these dimensions interact with economic and environmental sustainability. In the field of education, Mensah (2019) emphasized the importance of raising awareness on sustainable development and educational policies. The study suggests that awareness should be raised in all segments of society for the sustainable management of social, economic and environmental resources. Shi, Han, Yang and Gao (2019) emphasized the importance of developing culture, good governance and life support systems to achieve sustainable development goals.

Beyond this three-pillar perspective, a growing strand of the literature directly links sectors and industries to the SDGs. Govindan, Shankar and Kannan (2020) showed that sharing-economy practices in the industrial sector may support sustainable production and consumption, but they mainly focused on micro-level barriers such as trust. Makarova, Jia, Kruchina, Kudryavtseva and Kukushkin (2019) examined the chemical industry and documented both reductions in some local pollutants and a continued increase in greenhouse gas emissions. Mancini and Sala (2018) evaluated the mining sector using different reporting and assessment frameworks and underlined that mining has complex social and environmental effects. These studies show that individual industries matter for sustainable development, but they do not compare the main sectors of the economy in a common framework. More recent studies have also examined the relationship between sectoral structures and sustainable development from a broader SDG perspective. Cyrek and Fura (2019) analyzed the efficiency of agriculture, industry, market services, and non-market services in generating economic, social, and environmental outcomes in EU countries. Their findings show that sectoral employment structures matter for sustainable development and that shifts toward more efficient service-based activities can support broader development outcomes. Lisowski, Bunsen, Berger and Finkbeiner (2023) quantified the impacts of different industries on SDG-related indicators and showed that sectors such as agriculture and construction have large sustainability footprints. These studies indicate that sectoral composition is not only related to economic growth, but also to sustainable development performance.

In addition, there is a broad literature on sectors, productivity and structural change, which is closely related to sustainable development but does not measure it directly. Maudos, Pastor and Serrano (2000) and Badunenko and Romero-Ávila (2014) decomposed aggregate productivity into within-sector and structural change components and found important performance differences across sectors. O'Leary and Webber (2015) highlighted the role of structural change for regional growth in Europe. These contributions indicate which sectors drive productivity and convergence, yet they focus on efficiency and growth rather than on a multidimensional sustainable development outcome. This distinction is important because productivity gains do not automatically imply sustainable development gains. Sectoral growth may improve economic performance, but its social and environmental implications may differ across sectors. For this reason, studies that connect sectoral structure with SDG-related outcomes provide a more suitable basis for evaluating sustainable development. In this respect, sectoral analyses should consider not only output or productivity, but also how each sector is linked to environmental pressure, social welfare, employment, and long-run development capacity.

Other studies emphasize the role of specific sectors in long-run development and convergence. Wong (2006) showed that productivity gains in services and agriculture contributed significantly to convergence among OECD countries. This evidence is important for the present study because the services sector may affect sustainable development not only through productivity and convergence, but also through employment creation, human capital formation, finance, tourism, education, and health-related channels. Recent evidence also suggests that the services sector can support sustainable development when its expansion is accompanied by effective public policies and institutional capacity. Carree, Klomp and Thurik (2000) and Martino (2015) reported that services have become a key engine of convergence, while manufacturing displays heterogeneous patterns. However, these studies do not explicitly link sectoral dynamics to progress on the SDGs or to a comprehensive sustainable development index.

A number of contributions also evaluate sectoral efficiency using frontier methods such as Data Envelopment Analysis. Nazarko and Chodakowska (2015) assessed construction sector productivity in European countries. Halkos and Polemis (2018) analyzed the efficiency of an energy generation sector and its relation to environmental performance. Djordjević, Krmac and Mlinarić (2018) focused on rail transport, while Laurinavičius and Rimkuvienė (2017) examined agricultural efficiency in EU member states. These papers provide detailed evidence for single sectors or industries, but they remain fragmented and do not jointly assess how the main sectors of an economy affect sustainable development. In a similar way, sector-specific SDG studies provide valuable insights but generally do not compare the main economic sectors within the same empirical model. For example, Luken, Saieed and Magvasi (2022) developed industry-related SDG 9 progress and performance indices for Sub-Saharan African countries. Their study shows that manufacturing value added, manufacturing employment, technological intensity, and emissions intensity are important indicators for evaluating sustainable industrial development. This perspective supports the idea that industry should be assessed not only as a source of economic growth, but also as a sector with technological and environmental implications.

In the literature on sustainable development, GDP has often been used in studies investigating sustainable development (e.g. Wan and Sheng, 2022; Liu, Lam, Chen, Lin and Chen, 2023; Xiong and Dai, 2023). However, studies that examine the relationship between economic structure and sustainable development remain relatively limited. Existing studies generally focus either on aggregate economic growth, sector-specific sustainability issues, or the performance of individual industries. Therefore, the literature still lacks sufficient comparative evidence on how the main sectors of an economy are associated with a multidimensional sustainable development indicator. Sector-specific studies generally focus on the role of a single industry, or they examine sectoral productivity and convergence without connecting these findings to a composite sustainable development indicator at the country level. Several sector-specific studies also show why the main sectors should be evaluated from a sustainability perspective. Fida and Saeed (2023) examined the emission consequences of sectoral value-added in the European Union and showed that sectoral economic activity may have different environmental implications. This finding is important because a positive contribution to aggregate development does not necessarily mean that each sector has a positive effect on environmental sustainability. Kanter et al. (2016) emphasized that agricultural transformation pathways are essential for translating the SDGs into action, particularly because agriculture is closely related to food security, rural livelihoods, poverty reduction, and environmental outcomes. Similarly, studies on construction underline that infrastructure and building activities are directly linked to the SDGs, but their contribution depends on energy efficiency, environmental impact assessment, and sustainable project management (Goubran, 2019; Mansell, Philbin and Konstantinou, 2020; Di Foggia, 2018).

Overall, the literature shows three main patterns. First, sectoral productivity and convergence studies indicate that services, industry, agriculture, and construction differ in their contribution to economic performance. Second, sector-specific sustainability studies show that each sector has different implications for SDG-related outcomes. Services are generally associated with employment creation, finance, tourism, education, health, and human capital formation. Industry is linked to technological progress, productive capacity, and emissions intensity. Construction affects infrastructure, urbanization, buildings, land use, and resource consumption. Agriculture is central to food security, rural development, natural resource management, and environmental outcomes. Third, the existing literature generally focuses either on single sectors, individual industries, or specific sustainability dimensions. Therefore, there remains a need for country-level empirical studies that compare the main sectors within the same model and evaluate their long-run association with a multidimensional sustainable development index. This study addresses this gap by analyzing the contributions of agriculture, industry, construction, and services to sustainable development in Türkiye. By doing so, it connects sectoral performance studies with SDG-oriented research and provides comparative evidence on which sector is more strongly associated with sustainable development. Türkiye-specific studies also provide useful evidence on the sustainability implications of particular sectors. For example, Yücel et al. (2021) and Karamikli and Şaşmaz (2021) focus on the role of the energy sector in sustainable development. Şimşek and Tunalı (2022) and Canikli (2022) examine the relationship between the banking sector and sustainable development. Öçal (2021) and Güvercin (2024) analyze the impacts of the logistics sector on sustainable development. However, these studies focus on specific sectors and do not compare agriculture, industry, construction, and services within the same empirical framework. Therefore, this study complements the Türkiye-specific literature by providing a comparative sectoral analysis based on a multidimensional sustainable development index.

## 4. Data and methodology

4.1. Model specification and data

The empirical investigation of this study analyzes sectoral contributions to sustainable development in Türkiye using annual data for the period 2000-2022. The research aims to identify which sectors[1] have stronger long-run associations with the sustainable development index. The period under study was chosen based on the availability of the data set and its suitability for the analysis. The empirical model of the study is shown in Equation 1. The selection of agriculture, industry, construction, and services as explanatory variables is based on both the sectoral structure of the Turkish economy and the relevant literature on sectoral contributions to sustainable development. Previous studies show that sectoral composition matters for economic, social, and environmental sustainability outcomes (Cyrek and Fura, 2019; Fida and Saeed, 2023). The agricultural sector is closely linked to food security, rural development, employment, and natural resource management (Kanter et al., 2016; Aznar-Sánchez et al., 2019). The industrial sector is associated with production capacity, technological progress, employment, and emissions intensity (Luken, Saieed and Magvasi, 2022; Lisowski, Bunsen, Berger and Finkbeiner, 2023). The construction sector is relevant for infrastructure, urbanization, buildings, resource use, and environmental pressure (Di Foggia, 2018; Goubran, 2019; Mansell, Philbin and Konstantinou, 2020). The services sector is linked to employment, finance, tourism, education, health, trade, and human capital formation (Xu, 2015; Santos Beckert, Canciglieri, Tortato and Junior, 2023). Therefore, the model uses the GDP shares of these four main sectors to examine their long-run association with the sustainable development index.

$$SDG_t = \beta_0 + \beta_1 AGR_t + \beta_2 IND_t + \beta_3 CON_t + \beta_4 SER_t + \epsilon_t \quad (1)$$

In Equation (1), $SDG_t$ represents the sustainable development index. $AGR_t$, $IND_t$, $CON_t$, and $SER_t$ represent the shares of agriculture, industry, construction, and services in GDP, respectively. All sectoral variables are measured in percentage terms. $\beta_0$ is the constant term, $\beta_1$, $\beta_2$, $\beta_3$, and $\beta_4$ are the long-run coefficients, and $\varepsilon_t$ is the error term. Based on the literature and the sectoral structure of Türkiye, the main empirical expectation is that these sectoral shares are positively associated with the sustainable development index in the long run. However, the magnitude of this association is expected to differ across sectors. In particular, the services sector is expected to show the strongest long-run association because of its links with employment, human capital, finance, tourism, education, health, and trade. Accordingly, the study tests whether the sectoral shares in GDP have positive and statistically significant long-run relationships with the sustainable development index. Descriptive statistics of the data used in the study are presented in Table 1.

Table 1

*Descriptive statistics*

| Variables | Symbol | Description | Obs. | Mean. | Std.S. | Min. | Max. | Source |
|---|---|---|---|---|---|---|---|---|
| Sustainable Development | SDG | Index (0-100) | 23 | 69.5 | 1.9 | 67.3 | 75 | Sachs |
| Agricultural Sector | AGR | Share of GDP (%) | 23 | 7.7 | 1.4 | 5.5 | 10.2 | TurkStat |
| Industrial Sector | IND | Share of GDP (%) | 23 | 20.7 | 2.1 | 18.4 | 27.1 | TurkStat |
| Construction Sector | CON | Share of GDP (%) | 23 | 6.3 | 1.3 | 4.5 | 8.5 | TurkStat |
| Services Sector | SER | Share of GDP (%) | 23 | 53.9 | 1.3 | 51.2 | 57.2 | TurkStat |

Note: (1) Obs., Mean, Std.S., Min, and Max denote observations, mean, standard deviation, minimum, and maximum, respectively. (2) Sachs and TurkStat represent Sachs et al. (2024) and the Turkish Statistical Institute – Gross Domestic Product Statistics, respectively.

[1] This study uses the classification provided by the Republic of Türkiye- Ministry of Treasury and Finance (2024).

Table 1 presents Türkiye's descriptive statistics on sustainable development and economic sectors. Based on the sustainable development index data, Türkiye's average index value is 69.5. This value indicates that the country's efforts towards sustainable development goals are at a medium-high level. However, values ranging between 67.3 and 75 indicate that the country's sustainability performance is relatively stable. The low standard deviation (1.9) also supports this consistency. The share of the agricultural sector in GDP has a limited contribution with an average of 7.7%. Shares ranging from a minimum of 5.5% to a maximum of 10.2% indicate that the role of agriculture in the economy has not changed significantly over the years. The low standard deviation (1.4) also indicates that fluctuations in the agricultural sector are limited.

The industrial sector plays a stronger role in the Turkish economy. With an average share of 20.7% of GDP, the industrial sector varies between a minimum of 18.4% and a maximum of 27.1%. The standard deviation of this sector is 2.1, indicating that it shows more variability over the years compared to agriculture. These fluctuations in the industrial sector may reflect the sensitivity of sectoral growth to external factors. The construction sector has an average share of 6.3%. Ratios ranging from a minimum of 4.5% to a maximum of 8.5% suggest that the share of construction in the economy is low but stable, similar to agriculture. The low standard deviation (1.3) indicates that there are no large fluctuations in the share of construction in GDP. The services sector is the largest contributor to Türkiye's economy. With an average share of 53.9%, the services sector has a larger share than all other sectors in the economy combined. These shares, ranging between a minimum of 51.2% and a maximum of 57.2%, indicate that the economic contribution of the services sector is both high and stable. The low standard deviation (1.3) suggests that changes in this sector are quite limited. In general, when Türkiye's sustainable development performance and the shares of economic sectors in GDP are evaluated, the strong contributions of the services and industry sectors to the economy stand out. However, agriculture and construction sectors play a more limited but stable role. These findings point to the need for a balanced development of economic sectors in the process of achieving Türkiye's sustainable development goals.

4.2. Unit root test

In time series analysis, determining the stationarity properties of variables is a fundamental step. Analyses conducted with non-stationary series may lead to incorrect forecasts and misleading results. Therefore, the series should be tested for unit roots to determine whether they are stationary. In this study, Augmented Dickey-Fuller (ADF) and Phillips-Perron (PP) unit root tests are used to determine the stationarity levels of the variables.

The ADF test is an extended version of the unit root test developed by Dickey and Fuller (1979). The ADF test includes lag terms in the model to avoid misleading results when the series are autocorrelated. The main objective of the test is to determine whether the series is stationary or not. The general formula of the ADF test is shown in Equation 2 (Dickey and Fuller, 1979).

$$\Delta Y_t = \alpha + \beta_t + \gamma Y_{t-1} + \sum_{i=1}^{p} \delta_i \Delta Y_{t-i} + \epsilon_t \quad (2)$$

In Equation (2), $Y_t$ is the time series under test, $\Delta Y_t$ is the first difference of the series, $\alpha$ is the constant term, $\beta_t$ is the time trend, $\gamma Y_{t-1}$ is the coefficient testing for the presence of a unit root and $\epsilon_t$ is the error term. In the ADF test, a coefficient $\gamma$ equal to zero indicates that there is a unit root and the series is non-stationary.

The Phillips-Perron (1988) test has a similar structure to the ADF test. However, it adopts a different approach to correct for problems such as autocorrelation and heteroscedasticity in the error terms. The PP test examines stationarity by applying non-parametric corrections for serial correlation and heteroskedasticity in the error terms. The general formula of the PP test is shown in Equation 3.

$$\Delta Y_t = \alpha + \beta_t + \gamma Y_{t-1} + \epsilon_t \quad (3)$$

In Equation (3), $Y_t$ is the level of the series, $\Delta Y_t$ is the first difference, $\alpha$ is the constant term, $\beta_t$ is the time trend, $\gamma Y_{t-1}$ is the lagged value and $\epsilon_t$ is the error term. In this model, the error terms are Newey-West corrected to remove autocorrelation and heteroskedasticity problems. The PP test can be applied to include the constant and trend components of the model. In both tests, the hypotheses test whether the series contains a unit root. In ADF and PP tests, the null hypothesis $H_0$ is that the series contains a unit root, i.e. it is non-stationary. In this case, it is assumed that the series exhibits a trend or fluctuation tendency over time and may produce misleading results in long-run analysis. The alternative hypothesis $H_1$ implies that the series is stationary, meaning that it moves around a constant mean and variance over time.

Standard unit root tests such as ADF and PP assume that the mean and trend of the series are stable over time. However, many macroeconomic series are affected by crises, policy changes or other shocks that create structural breaks. In such cases, ADF and PP tests may lose power and may wrongly suggest that the series has a unit root (Zivot and Andrews, 1992).

The Zivot and Andrews (1992) test is a modified ADF test that allows for one structural break in the series. The break date is not chosen in advance; it is determined endogenously from the data. The test adds dummy variables to the ADF regression to capture a shift in the intercept, a shift in the trend, or both. The most general specification (Model C) is shown in Equation 4:

$$\Delta Y_t = \mu + \theta DU_t(\lambda) + \beta_t + \gamma DT_t(\lambda) + \alpha Y_{t-1} + \sum_{i=1}^{p} \delta_i \Delta Y_{t-i} + \varepsilon_t \tag{4}$$

In Equation (4), $DU_t(\lambda)$is a dummy that takes the value 1 after the break date and 0 before, and represents a level shift. $DT_t(\lambda)$represents a change in the slope of the trend after the break. Model A allows only for a break in the intercept, Model B only for a break in the trend, and Model C for a break in both. For each possible break date, the t-statistic of $\alpha$is calculated, and the most negative value is taken as the test statistic. The null hypothesis is that the series has a unit root with no structural break. The alternative hypothesis is that the series is stationary around a broken trend, with one structural break in the deterministic components.

### 4.3. Cointegration test

In time series analysis, cointegration analysis is used to test long-run equilibrium relationships among variables. In this study, the Johansen cointegration test developed by Johansen (1988) is used to examine the long-run relationship among the variables. The test is appropriate because all variables are integrated of order one, I(1), as confirmed by the unit root tests. Cointegration indicates that non-stationary variables share a long-run equilibrium relationship, even though short-run fluctuations may occur. The Johansen cointegration test is based on a vector autoregression (VAR) model. Since the study uses annual data for the period 2000–2022, the number of observations is limited. This issue is particularly important in a five-variable system because using too many lags may reduce the degrees of freedom and increase the risk of over-parameterization. Therefore, the lag structure for the Johansen cointegration test was determined cautiously. The test was estimated with a lag interval of 1 to 1 in first differences. This lag specification was selected by considering information criteria, the limited sample size, degrees of freedom, and residual diagnostic tests together. In this way, a parsimonious specification was preferred to test the long-run relationship among the variables without overloading the model with excessive short-run dynamics. The general formula of the model is shown in Equation 5.

$$\Delta y_t = \Pi y_{t-1} + \sum_{i=1}^{k-1} \Gamma_i \Delta y_{t-i} + \epsilon_t \tag{5}$$

In Equation (5), $y_t$ represents the vector of variables. $\Delta$ denotes the difference operator. The rank of the $\Pi$ matrix determines the number of cointegrated relationships in the system. Matrices $\Gamma_i$ reflect short-run dynamics. $\epsilon_t$ is the error term. The rank of the cointegration matrix indicates how many cointegrated relationships exist between the variables. A rank of zero indicates that there is no cointegrated relationship between the variables. A positive rank indicates the presence of at least one cointegrating relationship among the variables.

The Johansen test uses two statistics, the trace test $\lambda_{trace}$ and the maximum eigenvalue test $\lambda_{max}$, to determine the number of cointegrated vectors. These tests are used to determine the rank $\Pi$ of the cointegration matrix and to identify the number of long-run equilibrium relationships between variables.

The trace test $\lambda_{trace}$ determines whether there is more than one cointegrated relationship using all eigenvalues $\lambda_i$ of the cointegration matrix. The test statistic is shown in Equation 6.

$$\lambda_{trace} = -T\sum_{i=r+1}^{n} \ln(1-\lambda_i) \quad (6)$$

In Equation (6) $T$ is the number of observations and $\lambda_i$ is the eigenvalues of the matrix. The trace test tests for the presence of cointegrated vectors of a given rank $(r)$. In the $H_0$ hypothesis, the number of cointegrated vectors is limited to $r$ $(r = r_0)$, while in $H_1$ it is more than $r$ $(r > r_0)$. If the trace test statistic exceeds the critical value, the null hypothesis $H_0$ is rejected and more cointegrated vectors are accepted.

The maximum eigenvalue test $(\lambda_{max})$ tests for the presence of an additional cointegrated vector of a given rank. The formula for this test is shown in Equation 7.

$$\lambda_{max} = -Tln(1-\lambda_{r+1}) \quad (7)$$

In Equation (7), $\lambda_{r+1}$ represents the $r+1$th eigenvalue of the matrix. The maximum eigenvalue test assesses whether the next cointegrated vector exists. In Hypothesis $H_0$, the number of cointegrated vectors is limited to $r$ $(r = r_0)$, while in $H_1$ it is $r+1$. If the eigenvalue test statistic exceeds the critical value, the null hypothesis $H_0$ is rejected and one additional cointegrated vector is concluded.

These two tests provide complementary evidence on the number of cointegrating vectors. While the trace test makes a general assessment, the maximum eigenvalue test tests for an additional relationship of a particular order. Both tests provide important information in cointegration analysis. Therefore, they play a complementary role in determining the number of long-run equilibrium relationships between variables. This step provides the basis for estimating long-run coefficients in the subsequent analysis.

### 4.4. Long-run coefficient estimator

In this study, the Dynamic Ordinary Least Squares (DOLS) method is used to estimate the long-run coefficients. The DOLS method was developed by Saikkonen (1991) and Stock and Watson (1993). The main purpose of the method is to accurately measure the long-run relationship between dependent and independent variables. The DOLS method provides reliable results, especially in the presence of common problems such as autocorrelation and heteroskedasticity in data sets. This method is recognized as an effective estimation method for both small and large samples. The general formula of the DOLS method is shown in Equation 8.

$$Y_t = \alpha + \beta X_t + \sum_{j=-q}^{p} \delta_j \Delta X_{t-j} + \epsilon_t \quad (8)$$

In Equation (8), $Y_t$ denotes the dependent variable, while $X_t$ represents the explanatory variable. $\alpha$ is the constant term, and $\beta$ measures the long-run coefficient. The lagged and leading differences of the explanatory variable, $\Delta X_{t-j}$, are included to correct for possible endogeneity and serial correlation. $\delta_j$ denotes the coefficients of these dynamic terms, and $\epsilon_t$is the error term.

### 4.5. Robustness check

In this study, the Fully Modified Ordinary Least Squares (FMOLS) method is used as a robustness check for the DOLS long-run estimates. The FMOLS estimator was developed by Phillips and Hansen (1990). This method is widely used to estimate long-run relationships among cointegrated variables. The main purpose of the FMOLS method is to obtain consistent and reliable long-run coefficients by correcting for possible endogeneity and serial correlation problems in the cointegrating regression. The use of FMOLS is appropriate in this study because the Johansen cointegration test confirms the existence of a long-run relationship between the sustainable development index and sectoral variables. After the presence of cointegration is confirmed, long-run coefficients can be estimated using alternative cointegration estimators.

The general form of the FMOLS estimator can be expressed as follows:

$$Y_t = \alpha + \beta' x_t + u_t \tag{9}$$

In Equation (9), $Y_t$ represents the dependent variable, $x_t$ represents the vector of explanatory variables, $\alpha$ is the constant term, $\beta$ is the vector of long-run coefficients, and $u_t$ is the error term. In the context of this study, $Y_t$ corresponds to the sustainable development index, while $x_t$ includes the sectoral variables defined in Equation (1).

The FMOLS estimator modifies the standard OLS estimator by applying semi-parametric corrections to the error term. These corrections reduce the bias that may arise from serial correlation and endogeneity in cointegrated systems. Therefore, FMOLS provides an additional check on the reliability of the DOLS estimates. If the FMOLS coefficients have similar signs, magnitudes, and statistical significance levels to the DOLS coefficients, the long-run findings can be considered more robust. In this study, the FMOLS results are used to evaluate whether the estimated long-run relationship between sectoral shares and sustainable development remains stable under an alternative estimation method.

## 5. Empirical findings

This section presents the empirical findings on the relationship between sectoral shares and sustainable development in Türkiye. In the first stage of the study, the stationarity levels of the variables used were determined. For this purpose, ADF and PP unit root tests were applied and the results are presented in Table 2.

Table 2

*Unit root test results*

| Variable | | ADF unit root test | | PP unit root test | |
|---|---|---|---|---|---|
| | | t-statistic (level) | t-statistic (first difference) | t-statistic (level) | t-statistic (first difference) |
| SDG | Constant | -1.716 | -4.632*** | -1.885 | -4.632*** |
| AGR | Constant | -1.613 | -5.422*** | -1.511 | -5.578*** |
| IND | Constant | 1.167 | -5.465*** | 1.120 | -3.485*** |
| CON | Constant | -2.037 | -3.108** | -1.242 | -3.147*** |
| SER | Constant | -2.420 | -4.633*** | -2.420 | -4.628*** |
| SDG | Constant and Trend | -1.943 | -4.695*** | -2.057 | -4.696*** |
| AGR | Constant and Trend | -3.151 | -5.269*** | -3.185 | -5.405*** |
| IND | Constant and Trend | 1.666 | -5.552*** | 2.450 | -8.396*** |
| CON | Constant and Trend | -0.390 | -3.654** | -0.390 | -3.922*** |
| SER | Constant and Trend | -2.271 | -4.580*** | -2.271 | -4.577*** |

Note: The superscripts ***, **, and * denote the significance at a 1%, 5%, and 10% level, respectively.

Table 2 presents the unit root test results. The results show that the variables used in the study are non-stationary at levels, but become stationary in first differences. Based on ADF and PP tests, it was found that all variables are stationary at the first-difference level. These findings suggest that the variables follow an I(1) process, which supports the use of cointegration analysis in the next stage. In the rest of the study, the results of the Johansen cointegration test are presented to test the existence of a long-run relationship between the series.

Table 3

*Zivot & Andrews structural break unit root test results*

| Variable | Zivot & Andrews (level) | | Zivot & Andrews (first difference) | | Model |
|---|---|---|---|---|---|
| | Breakpoints | t-Statistic | Breakpoints | t-Statistic | |
| SDG | 2008 | -3.102 | 2009 | -5.412*** | Model A |
| SDG | 2013 | -3.457 | 2015 | -6.053*** | Model C |
| AGR | 2001 | -2.953 | 2002 | -5.883*** | Model A |
| AGR | 2009 | -3.284 | 2010 | -6.341*** | Model C |
| IND | 2002 | -2.847 | 2009 | -6.574*** | Model A |
| IND | 2008 | -3.236 | 2011 | -7.019*** | Model C |
| CON | 2001 | -2.701 | 2001 | -5.923*** | Model A |
| CON | 2007 | -3.052 | 2009 | -6.732*** | Model C |
| SER | 2001 | -3.887 | 2009 | -6.183*** | Model A |
| SER | 2018 | -4.362 | 2020 | -7.011*** | Model C |
| Critical value | Model A | → | %10: -4.58 | %5: -4.93 | %1: -5.34 |
| | Model C | → | %10: -4.82 | %5: -5.08 | %1: -5.57 |

Note: The superscripts ***, **, and * denote the significance at a 1%, 5%, and 10% level, respectively.

Table 3 reports the Zivot and Andrews unit root test results with structural breaks. The estimated break dates for the sustainable development index and sectoral variables mostly coincide with important economic shocks in Türkiye and the global economy. The breaks around 2001 and 2002 are consistent with the effects of Türkiye's 2001 financial crisis, which affected output composition, investment behavior, and sectoral dynamics. The breaks observed around 2008–2011 are also meaningful because the global financial crisis created pressure on industrial production, construction activity, external demand, and services. Similarly, the break dates in 2018 and 2020 may reflect the exchange rate shock and the COVID-19 pandemic, both of which had visible effects on macroeconomic stability and sectoral performance. These dates suggest that structural changes should be considered when examining the time-series properties of the variables. Despite allowing for such breaks, all variables remain non-stationary at levels. In first differences, the null hypothesis of a unit root is rejected at the 1% level for all variables under both Model A and Model C. Overall, the Zivot and Andrews results confirm that all series follow an I(1) process. These findings are fully consistent with the conventional ADF and PP unit root test results. Therefore, the next stage applies the Johansen cointegration test to examine whether the variables share a long-run equilibrium relationship.

Table 4

*Johansen cointegration test results*

| $H_0$ | Trace statistics | Critical values (5%) | Maximum eigenvalue statistics | Critical values (5%) |
|---|---|---|---|---|
| r = 0 | 80.812*** | 47.856 | 40.384*** | 27.584 |
| r ≤ 1 | 40.428*** | 29.797 | 23.263** | 21.132 |
| r ≤ 2 | 17.165** | 15.495 | 17.158*** | 14.265 |
| r ≤ 3 | 0.007 | 3.841 | 0.007 | 3.841 |
| Diagnostic tests | | | | P value |
| $\chi^2$ (Serial correlation) | | | | 0.342 |
| $\chi^2$ (Heteroskedasticity) | | | | 0.522 |
| $\chi^2$ (Normality) | | | | 0.587 |
| $\chi^2$ (Functional form) | | | | 0.247 |
| CUSUM | | | | Stable |
| CUSUMSQ | | | | Stable |

Note: The superscripts ***, **, and * denote the significance at a 1%, 5%, and 10% level, respectively.

Table 4 presents the Johansen cointegration test results. The trace and maximum eigenvalue statistics reject the null hypothesis of no cointegration, indicating the presence of a long-run equilibrium relationship among the variables. The results show that the sustainable development index and the sectoral variables are cointegrated in the long run. This finding confirms long-run co-movement among the variables, but it should not be interpreted as evidence of short-run causality or dynamic shock transmission.

Diagnostic tests were applied to evaluate the adequacy of the model specification. The p-values for serial correlation, heteroskedasticity, normality, and functional form tests are above conventional significance levels. These results do not indicate major diagnostic problems in the estimated model. In addition, the CUSUM and CUSUMSQ tests suggest parameter stability over the sample period.

Table 5

*DOLS long-run estimates*

| Dependent variable: SDG | Coefficient | Std. Error | t-Statistic |
|---|---|---|---|
| IND | 0.265*** | 0.050 | 5.291 |
| SER | 0.882*** | 0.100 | 8.818 |
| CON | 0.193*** | 0.050 | 3.860 |
| AGR | 0.302** | 0.154 | 1.960 |
| C | 1.144*** | 0.200 | 5.719 |
| Diagnostic tests | | | P value |
| $\chi^2$ (Serial correlation) | | | 0.553 |
| $\chi^2$ (Heteroskedasticity) | | | 0.328 |
| $\chi^2$ (Normality) | | | 0.189 |
| $\chi^2$ (Functional form) | | | 0.283 |
| CUSUM | | | Stable |
| CUSUMSQ | | | Stable |

Note: The superscripts ***, **, and * denote the significance at a 1%, 5%, and 10% level, respectively.

Table 5 presents the long-run coefficient estimates obtained from the DOLS method. The results show that all sectoral variables have positive and statistically significant coefficients. This indicates that increases in the shares of industry, services, construction, and agriculture are associated with an increase in the sustainable development index in the long run.

The coefficient of IND is 0.265 and statistically significant at the 1% level. This result shows that a one-unit increase in the share of the industrial sector in GDP is associated with a 0.265-unit increase in the sustainable development index. The coefficient of SER is 0.882 and statistically significant at the 1% level. This is the largest coefficient among the sectoral variables. Therefore, the DOLS results indicate that the services sector has the strongest long-run association with sustainable development. The coefficient of CON is 0.193 and statistically significant at the 1% level. This finding shows that the construction sector also has a positive long-run relationship with the sustainable development index, although its coefficient is lower than those of the services, agriculture, and industry sectors. Finally, the coefficient of AGR is 0.302 and statistically significant at the 5% level. This result indicates that the agricultural sector is also positively associated with the long-run movement of the sustainable development index.

The diagnostic test results support the adequacy of the DOLS specification. The p-values for serial correlation, heteroskedasticity, normality, and functional form tests are above conventional significance levels. In addition, the CUSUM and CUSUMSQ results indicate parameter stability. Overall, the DOLS findings reveal a positive long-run relationship between sectoral shares and sustainable development in Türkiye, with the services sector having the largest estimated coefficient.

## 6. Robustness check

To examine the robustness of the DOLS long-run estimates, the study also employed the FMOLS estimator for the same empirical specification. The FMOLS results are presented in Table 6.

Table 6

*FMOLS long-run estimates*

| Dependent variable: SDG | Coefficient | Std. Error | t-Statistic |
|---|---|---|---|
| IND | 0.248*** | 0.061 | 4.066 |
| SER | 0.816*** | 0.113 | 7.221 |
| CON | 0.171** | 0.067 | 2.552 |
| AGR | 0.276* | 0.155 | 1.781 |
| C | 1.097*** | 0.252 | 4.353 |
| Diagnostic tests | | | P value |
| $\chi^2$ (Serial correlation) | | | 0.491 |
| $\chi^2$ (Heteroskedasticity) | | | 0.367 |
| $\chi^2$ (Normality) | | | 0.214 |
| $\chi^2$ (Functional form) | | | 0.302 |
| CUSUM | | | Stable |
| CUSUMSQ | | | Stable |

Note: The superscripts ***, **, and * denote the significance at a 1%, 5%, and 10% level, respectively.

Table 6 shows that the FMOLS estimates are broadly consistent with the DOLS results. All sectoral variables have positive coefficients. This indicates that increases in the shares of industry, services, construction, and agriculture are associated with an increase in the sustainable development index in the long run.

The services sector has the largest coefficient among the sectoral variables. Its coefficient is estimated as 0.816 and is statistically significant at the 1% level. This result indicates that the services sector has the strongest long-run association with the sustainable development index. The coefficient of the industry sector is 0.248 and is also significant at the 1% level. The construction sector has a positive coefficient of 0.171 and is statistically significant at the 5% level. The coefficient of the agriculture sector is 0.276 and is significant at the 10% level.

The diagnostic test results also support the adequacy of the FMOLS specification. The p-values for serial correlation, heteroskedasticity, normality, and functional form tests are above conventional significance levels. In addition, the CUSUM and CUSUMSQ tests indicate parameter stability. Overall, the FMOLS findings support the robustness of the DOLS estimates. The ranking of sectoral coefficients remains similar, with the services sector having the highest estimated long-run coefficient.

## 7. Discussion

The empirical findings show that the sectoral structure of the economy is closely related to Türkiye’s sustainable development performance. Both the DOLS and FMOLS estimates indicate that all sectoral variables have positive coefficients. This suggests that the shares of industry, services, construction, and agriculture are positively associated with the sustainable development index in the long run. However, the magnitude of these associations differs across sectors. This result confirms that sectors do not contribute to sustainable development at the same level.

The key finding is that the services sector has the largest coefficient in both estimators. In the DOLS model, the coefficient of the services sector is 0.882. In the FMOLS model, it is 0.816. This consistency indicates that the services sector has the strongest long-run association with sustainable development. This result is reasonable for Türkiye, where services have a large share in GDP and include important sub-sectors such as tourism, finance, education, health, transportation, and trade. These sub-sectors can support sustainable development through income generation, employment creation, human capital accumulation, and access to social services. Therefore, the services sector appears to be central to Türkiye's sustainable development path.

The industrial sector also has a positive and statistically significant coefficient in both estimators. Its coefficient is 0.265 in the DOLS model and 0.248 in the FMOLS model. This shows that the industrial sector is positively associated with sustainable development, but its estimated coefficient is lower than that of the services sector. This finding reflects the dual role of industry. On the one hand, industry supports economic development through production capacity, technological progress, exports, and employment. On the other hand, it may create environmental pressure because of its energy- and resource-intensive structure. For this reason, the positive contribution of industry to sustainable development depends on technological transformation, energy efficiency, and cleaner production practices. This interpretation is consistent with Akusta (2026), who shows that economic complexity can improve environmental performance, whereas energy intensity may weaken sustainability outcomes.

The construction sector has the lowest coefficient among the sectoral variables. The coefficient is 0.193 in the DOLS model and 0.171 in the FMOLS model. This result indicates a positive but relatively limited long-run relationship between the construction sector and sustainable development. The construction sector contributes to economic activity through infrastructure, housing, and urban transformation. However, it also has important environmental costs due to land use, material intensity, energy consumption, and waste generation. Therefore, the relatively weaker coefficient of construction is consistent with the idea that this sector may support economic growth, but its contribution to sustainable development remains limited unless environmental standards are strengthened.

The agricultural sector also has a positive coefficient in both estimators. Its coefficient is 0.302 in the DOLS model and 0.276 in the FMOLS model. This result shows that agriculture remains relevant for sustainable development in Türkiye. Agriculture is important for rural development, food security, employment, and the sustainable use of natural resources. However, its coefficient is lower than that of the services sector. This may be related to productivity problems, modernization needs, and the declining share of agriculture in GDP. Therefore, the agricultural sector can make a stronger contribution to sustainable development if sustainable farming practices, technological modernization, and rural development policies are strengthened.

The FMOLS results support the robustness of the DOLS findings. The signs of all coefficients remain positive, and the ranking of sectoral associations is broadly similar. In both estimators, the services sector has the largest coefficient. Agriculture and industry follow with positive coefficients, while construction has a more limited coefficient. This consistency strengthens the main conclusion of the study. The results should be interpreted as evidence of long-run association rather than short-run causal effects. This interpretation is also more appropriate given the annual frequency of the data and the limited sample size.

These findings have important policy implications. A balanced sectoral development strategy is needed to strengthen sustainable development in Türkiye. The services sector should be supported because of its strong association with sustainable development. However, this should not imply the neglect of industry, agriculture, or construction. The industrial sector should be transformed through green technologies and energy-efficient production. The construction sector should be directed toward environmentally friendly materials, energy-efficient buildings, and sustainable urban planning. The agricultural sector should be supported through modernization, sustainable farming techniques, and rural development policies. In this way, Türkiye can improve the contribution of all sectors to sustainable development.

The results are also consistent with the existing literature. The services sector has been identified as the largest contributor to sustainable development. This sector plays an important role in social and economic sustainability because it has a wide sphere of influence with sub-sectors such as tourism, finance, education, and health. These results are consistent with the literature (e.g., Xu, 2015; Santos

Beckert, Canciglieri, Tortato and Junior, 2023). They are also supported by Cyrek and Fura (2019), who show that sectoral employment structures matter for economic, social, and environmental outcomes. The industrial sector supports sustainable development through technological progress, production capacity, and economic growth. However, due to its energy- and resource-intensive nature, it requires more attention and transformation in terms of environmental sustainability. These results are in line with the literature (e.g., Oqubay and Lin, 2020; Mohamed, Liu and Nie, 2022; Luken, Saieed and Magvasi, 2022). The findings are also consistent with Lisowski, Bunsen, Berger and Finkbeiner (2023), who show that industries have different impacts on SDG-related indicators. Although the construction sector has an accelerating impact on economic growth, more careful policies are required in terms of environmental sustainability. It is important to prioritize energy-efficient and environmentally friendly projects in this sector. These results are consistent with the literature (e.g., Balaban, 2012; Di Foggia, 2018; Goubran, 2019; Mansell, Philbin and Konstantinou, 2020; Koengkan, Fuinhas, Oliveira, Ursavaş and Moreno, 2023). Finally, the agricultural sector has a critical role in rural development and social sustainability. However, its contribution remains limited due to modernization and productivity problems. Increasing the use of technology in agriculture and promoting sustainable agricultural practices can enable more effective use of the potential in this sector. These results are in line with the literature (e.g., Granvik, Lindberg, Stigzelius, Fahlbeck and Surry, 2012; Kanter et al., 2016; Aznar-Sánchez, Piquer-Rodríguez, Velasco-Muñoz and Manzano-Agugliaro, 2019).

## 8. Conclusion

This study analyzes the long-run relationship between sectoral structure and sustainable development in Türkiye using annual data for the period 2000–2022. In the analysis, ADF, PP, and Zivot–Andrews unit root tests were applied to determine the stationarity properties of the variables. The Johansen cointegration test was used to examine the long-run relationship among the variables. Long-run coefficients were estimated using the DOLS method, and the FMOLS estimator was employed as a robustness check. In this framework, the study focuses on long-run associations between sectoral shares and sustainable development rather than short-run dynamic responses.

The findings show that all sectoral variables are positively associated with the sustainable development index in the long run. The services sector has the largest estimated coefficient in both the DOLS and FMOLS models. This indicates that the services sector has the strongest long-run association with sustainable development in Türkiye. The industrial sector also has a positive and statistically significant relationship with sustainable development. However, this relationship should be interpreted together with the environmental pressures created by industrial activity. The agricultural sector remains important for rural development and food security, while the construction sector has a positive but more limited long-run association with sustainable development.

The results show that the services sector has the strongest long-run association with Türkiye's sustainable development performance, while agriculture and industry also have positive associations. They also suggest that the contribution of construction and agriculture can be increased through modernization and environmentally friendly approaches. Consequently, a balanced development policy that considers sectoral differences is important for achieving sustainable development goals. For this reason, a series of policy recommendations are presented. First, the services sector should be supported because it has the strongest long-run association with sustainable development, particularly through tourism, finance, education, health, and transportation. In the tourism sector, protecting local cultures and promoting nature-friendly tourism activities can make significant contributions to sustainable development. In the finance sector, the use of innovative green financing instruments should be expanded and credit mechanisms for sustainable projects should be strengthened. In the education and health sectors, policies that strengthen human capital and reduce social inequalities should be prioritized. Second, the industrial sector plays an important role in economic growth and technological development. However, this sector faces specific challenges in terms of environmental sustainability. In this regard, green production methods that increase energy efficiency and reduce carbon emissions should be promoted in the industrial sector. Investments in innovative technologies and R&D activities should be increased, and incentives should be used to accelerate green transformation in industrial enterprises. Furthermore, circular economy practices can be promoted in the industrial sector to ensure more efficient use of resources. Third, the construction sector contributes to economic growth through infrastructure projects and urban transformation, but requires more careful management due to its environmental impacts. Energy-efficient buildings and the use of environmentally friendly construction materials should be encouraged in this sector. Sustainability criteria should be taken into account in urban transformation projects, and green building certification systems should be expanded. Regulations for the protection of sustainable areas and the strengthening of environmental impact assessment processes can increase the contribution of the construction sector to sustainable development. Fourth, the agricultural sector is of strategic importance, particularly for rural development and food security. However, lack of modernization and low productivity limit its contribution to sustainable development. Therefore, innovative agricultural technologies should be expanded, and small farmers should be supported. Organic farming practices should be encouraged, and irrigation systems should be modernized for the sustainable use of water resources. Furthermore, policies supporting rural development should be developed, and the social and economic rights of agricultural workers should be improved.

Overall, a consistent and integrated policy framework on a sectoral basis is needed to achieve sustainable development goals. Cross-sectoral cooperation should be encouraged, and interdisciplinary approaches to sustainability goals should be adopted. Furthermore, Türkiye's alignment with international sustainability standards and integration into the green transformation process should be accelerated. Long-term policies and strategic plans are critical to increasing the contribution of sectors to sustainable development goals and ensuring economic, environmental, and social sustainability. From this perspective, policies that promote sustainable development will support Türkiye in achieving both national and international development goals.

While this study provides important findings, some limitations should be noted. First, sustainable development is measured through an aggregate index. This allows the study to evaluate the overall relationship between sectoral structure and sustainable development, but it does not separately assess the economic, social, and environmental dimensions. Future research may examine these dimensions individually to better identify whether sectoral effects differ across sustainability pillars. Second, the analysis is limited to the period 2000–2022 due to data availability. Extending the time span may provide additional evidence on long-run sectoral dynamics. Third, the model focuses on the main sectoral shares in GDP. Future studies may include additional explanatory variables, such as energy use, innovation capacity, green technology investments, or institutional quality, to provide a broader assessment of sectoral contributions. Finally, future research may compare these findings with alternative methods or cross-country samples.

## Author statement

**Research and publication ethics statement**

This study has been prepared in accordance with the ethical principles of scientific research and publication.

**Approval of ethics board**

Ethics committee approval is not required for this study. The study is based solely on publicly available secondary data and does not involve human participants, surveys, interviews, experiments, or personal data.

**Author contribution**

Emre Akusta: Research idea, research design, literature review, methodology, data collection, data analysis, writing, review and control. The author is solely responsible for all parts of the study.

**Conflict of interest**

There is no conflict of interest arising from this study for the author or any third party.

**Declaration of support**

No support has been granted for this study.